\documentclass[a4paper]{article}

\usepackage{mathptmx}
\usepackage{amsmath}
\usepackage{graphicx}
\usepackage[utf8]{inputenc} 

\providecommand{\keywords}[1]
{
  \small	
  \textbf{Keywords: } #1
}

\begin{document}

\title{Multiple imputation using chained random forests: a preliminary study
based on the empirical distribution of out-of-bag prediction errors}

\author{Shangzhi Hong$^{1}$, Yuqi Sun$^{1}$, Hanying Li$^{1}$, 
Henry S. Lynn$^{1,*}$\\
\small $^{1}$Department of Biostatistics, School of Public Health, Fudan
University}

\maketitle

\begin{abstract}
  Missing data are common in data analyses in biomedical fields, and imputation
  methods based on random forests (RF) have become widely accepted, as the RF
  algorithm can achieve high accuracy without the need for specification of data
  distributions or relationships. However, the predictions from RF do not
  contain information about prediction uncertainty, which was unacceptable for
  multiple imputation. Available RF-based multiple imputation methods tried to
  do proper multiple imputation either by sampling directly from observations
  under predicting nodes without accounting for the prediction error or by
  making normality assumption about the prediction error distribution. In this
  study, a novel RF-based multiple imputation method was proposed by
  constructing conditional distributions the empirical distribution of
  out-of-bag prediction errors. The proposed method was compared with previous
  method with parametric assumptions about RF’s prediction errors and predictive
  mean matching based on simulation studies on data with presence of interaction
  term. The proposed non-parametric method can deliver valid multiple imputation
  results. The accompanying R package for this study is publicly available.
\end{abstract}


%
%
%
%
\section{Introduction}
\label{sec:1}
Missing data have been a common nuisance in data analyses. To handle missing
data problems, different imputation methods was proposed. For the past few
years, with the fast development of machine learning methods, imputation methods
based on machine learning algorithms have been proposed. Such imputation
methods, specially imputation methods based on random forest (RF)
\cite{Breiman2001}, have drawn attentions from statisticians \cite{VanBuuren2018}
because of their abilities to handle data without the need to specify the
distributions of the variables like most standard methods, and can handle
complex relationships in the datasets automatically \cite{Biau2016} with high
imputation accuracy \cite{Stekhoven2011}.

However, such methods have incompatibility with traditional statistical methods.
One major problem is that the predictions made by the RF algorithm does not
contain information about the uncertainty of the predictions, which is
unacceptable for multiple imputation. Using predictions from RF for imputation
can cause less variability of imputations, narrower width of confidence
intervals, and under coverage of confidence intervals \cite{Shah2014} . To get
“proper” multiple imputation results, researchers have tried to take the
prediction uncertainty of RF into account and construct conditional
distributions based on RF predations. Shah \textit{et al.} \cite{Shah2014} assumed
normality for RF prediction errors and used RF’s out-of-bag (OOB) mean squared
error (MSE) as the estimate for the prediction error variance, while OOB MSE can
be severely biased for estimated the variance for prediction errors
\cite{Bylander2002, Mitchell2011}. Doove \textit{et al.} \cite{Doove2014} sampled
directly from observed values under the predicting nodes of RF without
accounting for the prediction errors of RF.

Researchers have proposed different methods for describing the prediction
uncertainty of RF, such as quantile regression forest by Meinshausen
\cite{Meinshausen2006}, jackknife and infinitesimal jackknife by Wager 
\textit{et al.} \cite{Wager2014}, split conformal by Lei \textit{et al.} 
\cite{Lei2017}. Recently, Zhang \textit{et al.} \cite{Zhang2019} used OOB
prediction errors to form empirical error distribution of RF prediction for the
construction of prediction intervals of RF, and this method can achieve valid
results under certain conditions without additional repetitive computations.
However, this empirical OOB prediction error distribution method relied on the
number of trees grown in the RF model (less number of trees grown can lead to
observations not out-of-bag for all the trees thus less valid observations in
the empirical distribution), while for RF-based multiple imputation, often only
a few trees (less than 20) were recommended by Shah \textit{et al.}
\cite{Shah2014} It is not clear whether the empirical OOB error distribution can
be used for constructing conditional distributions in multiple imputation.

In this paper, a novel RF-based multiple imputation method was proposed based on
the empirical error distribution of RF and compared with previously established
multiple imputation method with normality assumption for RF prediction errors,
as well as prediction mean matching (PMM), for data with presence of interaction
term.

%
%
%
%
\section{Methods}
\label{sec:2}
\subsection{Empirical distribution of out-of-bag prediction errors}
\label{sec:2.1}
The original RF algorithm was proposed by Breiman \cite{Breiman2001}, and
afterwards Liaw \textit{et al.} \cite{Liaw2002} provided an easy-to-use
implementation in R named “randomForest”. As an ensemble learning method, RF
used bootstrap aggregation (bagging) \cite{Breiman1996} to reduce the risk of
overfitting and produce accurate predictions based on predictions from a number
of random trees. For the process of constructing a random tree, the training set
used was an independent bootstrap sample of the input dataset, and the
unselected observations for a certain random tree is called the OOB sample for
the tree. For a certain observation, predictions from the subset of RF that the
observation is OOB, is called the OOB prediction, and the difference of input
value and the OOB prediction is the OOB error. The empirical error distribution
is from the individual OOB errors. The MSE of the OOB errors was used by Shah et
al. \cite{Shah2014} as the estimate of RF prediction error variance.

\subsection{Multiple imputation using chained forests}
\label{sec:2.2}
Using the framework of MICE (multivariate imputation using chained equations),
multiple imputation using chained forests is based on drawing data samples from
conditional distributions constructed using RF, which can be summarized as:

First, initializing the simulation chains: the number of simulation chains
equals to the number of imputations, and the missing part of the variable is
replaced by random samples of the observed values of the variable.

Second, constructing conditional distributions based on RF: two distinct parts
are formed based on whether the variable is missing, and the observed part is
used as the input for training set of RF, and the missing part is used as the
input for prediction set of RF. The bootstrap example of the training set is
achieved from the input to account for the sampling variation. And the
predictions from the prediction set was used to get the RF predictions. For
continuous variables, the prediction errors of RF were account for by the
empirical distribution of out-of-bag prediction errors. For categorical
variables, the classes for predictions were assigned randomly according to the
predicted probabilities.

For each of the simulation chains, the second step is performed iteratively
multiple times, and the results from the last iteration were used. The software
package “RfEmpImp” is now publicly available on GitHub \cite{Hong2020}.

\subsection{Simulation studies}
\label{sec:2.2}
A series of simulations and analyses were carried out using R, version 3.6 (R
Core Team, Vienna, Austria). Four sequential stages were involved:
\begin{enumerate}
  \item Data generation: complete datasets were simulated based on pre-defined
scenarios.
  \item Amputation: the complete datasets were made incomplete based on
specified rules.
  \item Imputation: the missing values contained in the simulated datasets were
filled in by missForest using different parallel strategies.
  \item Analysis: Statistical analysis were performed on both the original
complete datasets and the corresponding imputed datasets, and comparisons
were made.
\end{enumerate}

\subsubsection{Data generation}
\label{sec:2.2.1}
Altogether, a total 1000 simulated datasets containing 2000 observations each
were generated based on following settings.
$$X \sim \text { Normal }(2,1)$$
$$Z \sim \text { Normal }(2,1)$$
$$Y=X-XZ+Z+\varepsilon, \varepsilon \sim \text { Normal }(0,1)$$

\subsubsection{Amputation}
\label{sec:2.2.2}
Missing data mechanisms can be classified into three categories
\cite{Rubin1976}. When data are MCAR, the probability of being missing is the
same for all cases. When data are MAR, the probability of being missing is only
related to (some of) the observed data. If neither MCAR nor MAR holds, then data
are missing not at random (MNAR). While MCAR is simple to consider, most of the
missing data methods use MAR assumption. Amputation functions provided by the
“mice” R package \cite{VanBuuren2011, Schouten2018} were used in this study to
generate missing values. Both missing completely at random (MCAR) and missing at
random (MAR) patterns were used in this study. MCAR patterns were introduced by
setting $X$ and $XZ$ to be missing with probability of each observation being
missing was set to 50\%. MAR patterns were introduced by setting $X$ and $XZ$ to
be missing depending on Y. Specifically, the probability of each observation
being missing was set to 50\% according to a standard right-tailed logistic
function on $Y$; thus the probability of the covariates being missing is higher
for observations with higher values of $Y$.

\subsubsection{Imputation}
\label{sec:2.2.3}
Multiple imputation was performed using “mice” \cite{VanBuuren2011} R package.
For each of the amputed dataset, RF-based imputation using empirical OOB error
distribution (“Empirical”) , RF-based imputation using normality assumption
(“Normal”), and PMM were performed. Ten imputations were performed for each
dataset and each imputation method, and the number of iterations restricted to
10 \cite{Shah2014}. The PMM method is a semi-parametric imputation method
recommended as the default method for handling missing data in continuous
variables by the “mice” R package. For each variable, PMM calculates the
predicted regression values for its non-missing and missing observations. It
then fills in a missing value by randomly selecting one from the “donors”
(non-missing observations whose predicted values are closest to the predicted
value for the missing observation). The purpose of the regression in PMM is to
construct a metric for matching observations with missing values to similar
observations with observed values that can be used for imputation.

\subsubsection{Analysis}
\label{sec:2.2.4}
Performance comparisons were made among the two RF-based imputation methods
(“Empirical” and “Normal”), PMM imputation, and results from corresponding
original data (“Original”), based on following statistics \cite{Vink2016}:
Comparisons were made between the two parallel strategies, along with the
original sequential algorithm, based on:

\begin{enumerate}
  \item the relative bias of the coefficient estimate:
    $$
      \frac{\left(\widehat{\beta}_{p}-\beta_{p}\right)}{\beta_{p}},
      p = 1 \text{ or } 2
    $$
    corresponding to the intercept (if any), $X_{1}$ or $X_{2}$.
  \item width of 95\% confidence intervals (CIs);
  \item coverage of 95\% CIs.
\end{enumerate}
An imputation method with superior performance can be generally characterized by
smaller relative bias, width of 95\% CIs closer to original data, coverage of
CIs closer to 95\%.

%
%
%
%
\section{Results}
\label{sec:3}
\subsection{Linear regression with interaction term}
\label{sec:3.1}
\begin{figure}[h!]
  \includegraphics[width=\linewidth]{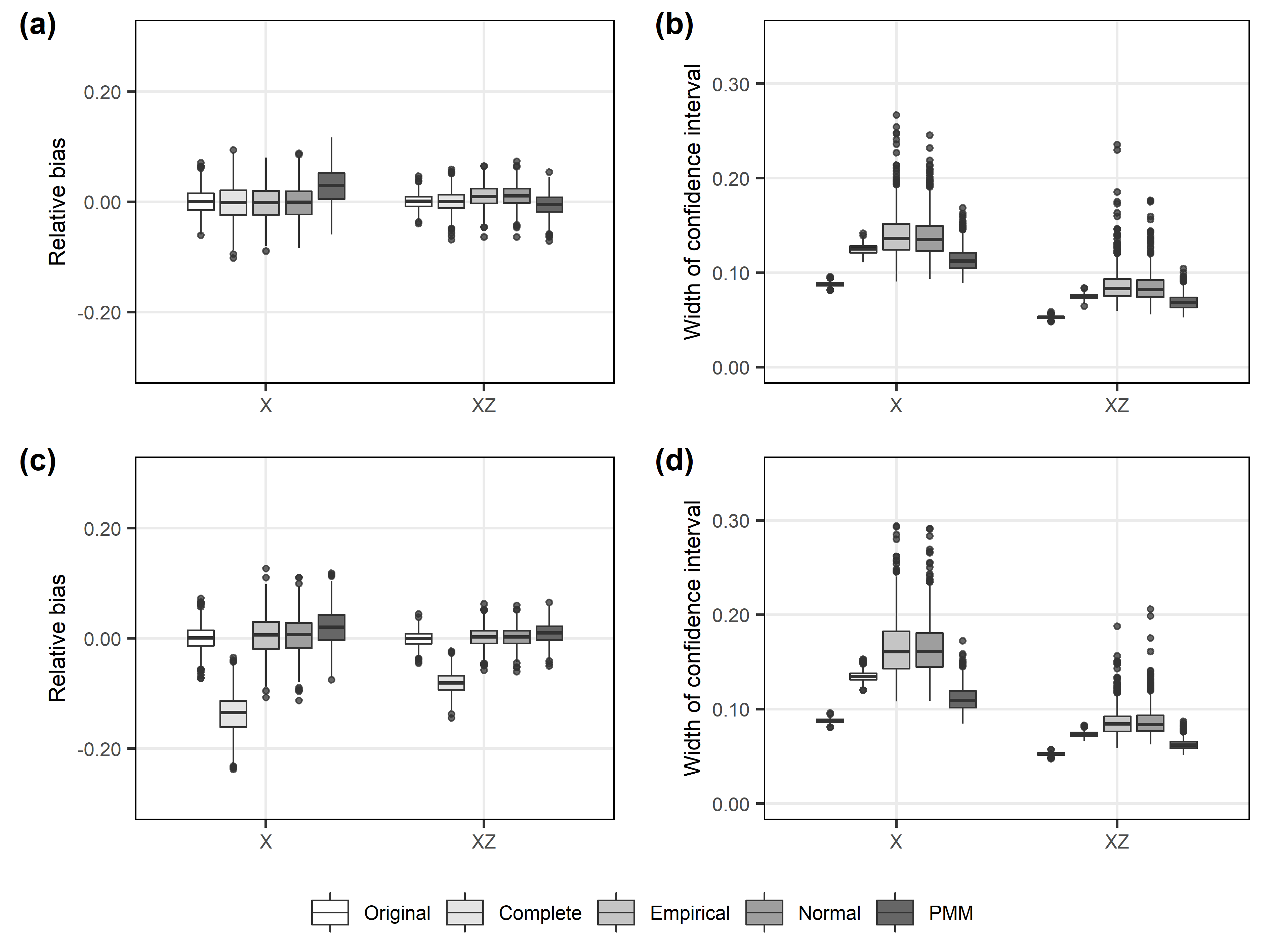}
  \caption{Relative bias of coefficient estimates of $X$ and $XZ$ when data
  were (a) MCAR or (c) MAR; width of confidence intervals of coefficient 
  estimates of $X$ and $XZ$ when data were (b) MCAR or (d) MAR for linear 
  regression with interaction of positive coefficient for $X$ and negative
  coefficient for $XZ$.}
  \label{fig:1}
\end{figure}

(1) MCAR data with positive coefficient for interaction and negative coefficient
for $X$ (Fig. 1ab). For relative bias of coefficient estimates of $X$, PMM
imputation can lead to upward bias (median=0.1\%, -0.1\%, -0.1\%, -0.1\%, 3.0\%,
for“Original”, “Complete”, “Empirical”, “Normal”, and “PMM”, respectively). For
CI width of $X$'s coefficients estimates, all methods can lead to increased CI
width (median=0.09, 0.12, 0.14, 0.14, 0.11). For CI coverage of $X$’s
coefficients estimates, PMM can lead to less coverage rate than other methods
(percent=95.0\%, 95.5\%, 97.6\%, 97.0\%, 78.9\%) For relative bias of
coefficient estimates of $XZ$, overall, the bias was small (median=0.1\%, 0.1\%,
1.0\%, 1.1\%, -0.5\%). For CI width of $XZ$'s coefficients estimates, RF-based
methods can lead to increased CI width (median=0.05, 0.07, 0.08, 0.08, 0.07).
For CI coverage of $XZ$’s coefficients estimates, all methods can lead to
reasonable CI coverage, and RF-based methods can have higher CI coverage
(percent=94.9\%, 95.7\%, 94.7\%, 93.7\%, 92.0\%).

(2) MAR data with positive coefficient for interaction and negative coefficient
for $X$ (Fig. 1cd). For relative bias of coefficient estimates of $X$,
complete-case analysis can lead to downward biased results, while RF-based
methods can lead to slightly upward biased results (median=0.0\%, -13.5\%,
0.6\%, 0.7\%, 2.0\%). For CI width of $X$'s coefficients estimates, RF-based
methods can lead to increased CI width (median=0.09, 0.13, 0.16, 0.16, 0.11).
For CI coverage of $X$’s coefficients estimates, complete-case analysis can lead
to poor CI coverage, and RF-based methods can lead to high CI coverage, higher
than that from PMM imputation (percent=94.7\%, 2.6\%, 97.3\%, 97.7\%, 83.6\%).
For relative bias of coefficient estimates of $XZ$, RF-based methods can lead to
nearly unbiased results, while PMM imputation can lead to slight upward bias,
and complete-case analysis can lead to downward bias(median=0.0\%, -8.1\%,
0.3\%, 0.3\%, 1.0\%). For CI width of $XZ$'s coefficients estimates, RF-based
methods can lead to increased CI width (median=0.05, 0.07, 0.08, 0.08, 0.06).
For CI coverage of $XZ$’s coefficients estimates, RF-based methods can lead to
high CI coverage, PMM imputation can lead to under-coverage, while CI coverage
from complete-case analysis was near zero (percent=95.6\%, 1.0\%, 97.8\%,
98.5\%, 86.7\%).

%
%
%
%
\section{Discussion}
\label{sec:4}
In this study, a novel RF-based imputation method was proposed and compared with
existing methods, results showed that the proposed method can produce valid
multiple imputation results and can provide less biased results for certain
scenarios. So, the normality assumption about RF’s prediction errors can be
relaxed.

Compared with existing RF-based multiple imputation methods, the proposed method
did not make parametric assumptions about the RF prediction errors. As a
data-driven method, the prediction errors of RF can be also data-dependent, so
strong parametric assumptions about RF prediction errors may not be valid. Also,
the OOB MSE has been reported to be a biased estimate of the variance of RF
prediction errors. The simulated scenarios in this study is somewhat simplistic
for real-word analyses, and the number of trees (10 trees in RF) used was small
to avoid over-fitting when training RF models. However, as the empirical
distribution of OOB prediction errors depends on the number of trees
constructed, the parameter of tree numbers may need further investigation for
practical implementation for multiple imputation. In the study, the RF model
building process was accelerated using parallel computing \cite{Wright2017}
(time consumption of RF-based methods is about 1.4x of PMM for simulated
datasets in this study), however, whether changes can be caused by different RF
software packages may need further discussion.

In this study, the simulations generated a relatively high proportion of missing
data to accentuate the effects of missing data on the results, although lowering
the proportion of missing data to 25\% did not materially alter the main
findings, while the complete-case analysis can produce slightly less biased
results. In order to be consistent with previous studies, we used only 10
imputations with 10 iterations for comparisons, but for practical use, more
stable pooled estimates may be achieved with more imputations and iterations.

%
%
%
%
\section{Conclusions}
\label{sec:5}
The proposed RF-based multiple imputation method based on the empirical
distribution of out-of-bag prediction errors can provide valid imputation
results even with only a small number of trees built in the model.

\bibliographystyle{unsrt}
\bibliography{Manuscript3_ShangzhiHong_Main}{}

\end{document}